\documentclass[nofootinbib,preprint,showpacs,preprintnumbers,amsmath,amssymb,aps,pre,showkeys]{revtex4-2}

\usepackage{graphicx} 
\usepackage[english]{babel}
\usepackage{amssymb}
\usepackage{xcolor}
 \usepackage{amsmath} 
\usepackage{etoolbox}
\usepackage{array} 
\usepackage{ragged2e} 

\usepackage{adjustbox}    
\usepackage{setspace}

\begin{document}

\title{Two-Player Yorke's Game of Survival in Chaotic Transients}

\author{Gaspar Alfaro}
\affiliation{Nonlinear Dynamics, Chaos and Complex Systems Group, Departamento de  Física, Universidad Rey Juan Carlos, Tulipán s/n, Móstoles, 28933, Madrid, Spain}

\author{Ruben Capeáns}
\affiliation{Nonlinear Dynamics, Chaos and Complex Systems Group, Departamento de  Física, Universidad Rey Juan Carlos, Tulipán s/n, Móstoles, 28933, Madrid, Spain}

\author{Miguel A.F. Sanjuán}
\affiliation{Nonlinear Dynamics, Chaos and Complex Systems Group, Departamento de  Física, Universidad Rey Juan Carlos, Tulipán s/n, Móstoles, 28933, Madrid, Spain}

\date{\today}

\begin{abstract}

We present a novel two-player game in a chaotic dynamical system where players have opposing objectives regarding the system's behavior. The game is analyzed using a methodology from the field of chaos control known as partial control. Our aim is to introduce the utility of this methodology in the scope of game theory. These algorithms enable players to devise winning strategies even when they lack complete information about their opponent's actions.  To illustrate the approach, we apply it to a chaotic system, the logistic map. In this scenario, one player aims to maintain the system's trajectory within a transient chaotic region, while the opposing player seeks to expel the trajectory from this region. The methodology identifies the set of initial conditions that guarantee victory for each player, referred to as the \textit{winning sets}, along with the corresponding strategies required to achieve their respective objectives. 

\end{abstract}


\maketitle
\newpage

\section{Introduction}

Game theory provides powerful tools for analyzing strategic interactions across diverse fields, from social sciences to economics and physics \cite{Social,EconomyGames,GamesComplex}. While classical game theory typically focuses on equilibrium states, many real-world situations involve chaotic systems, where extreme sensitivity to initial conditions and inherent unpredictability create fundamental challenges for strategic decision-making. The nonlinear nature of these systems makes traditional game-theoretic approaches insufficient, as small perturbations can lead to dramatically different outcomes. What's more, the intrinsic characteristics of a game are susceptible to change in real life, and these changes can result from the players' decisions \cite{AkiyamaKaneko1,AkiyamaKaneko2}.

Game theory has also proven valuable in chaos control \cite{GamesControl}, where control problems naturally emerge as competitive scenarios between opposing objectives. This framework reveals how controllers must optimize their strategies while dealing with three key challenges: (1) the unpredictable nature of chaotic dynamics, (2) the system constraints, and (3) the actions of other controllers who must carefully choose their actions to achieve their own goal. This interaction between different control agents adds a strategic dimension that goes beyond traditional chaos control methods.

In this paper, we introduce a competitive control game in a scenario governed by chaotic dynamics, where two players with limited control capabilities aim to steer trajectories toward opposing objectives.  One player seeks to confine the system within a target set $Q$, while the other attempts to drive it out. This framework builds on the approach introduced in \cite{Yorke}, where the authors propose a control game in a transient chaotic region.

Two crucial aspects that determine the victory are the initial state of the system and the information that each player has about their opponent's moves. The starting point determines which player has a natural advantage due to the underlying dynamics. On the other hand, with information about opponent moves, a player can better choose its actions and adopt a control accordingly. In this regard, we explore three different scenarios: (1) whether one player knows the actions of the opponent, e.g., by moving after it, (2) whether the other player knows the actions, or (3) whether neither player has such knowledge, they move simultaneously.

Our main contribution is to develop a method to solve the game. This method determines, for any initial condition in phase space, when and where a player can guarantee victory regardless of the opponent actions, based solely on the control bounds and the system dynamics. 

To this end, we develop a methodology involving the computation of \textit{winning sets} corresponding to the \textit{safe sets} previously introduced in the context of controlling transient chaos \cite{DynamicsPartialControl,PartialControlBeyond,PartialControlFunctions,PartialControlEscape}. To illustrate our approach, we use the logistic map in its chaotic regime as a paradigmatic example. Our analysis reveals cases where only one player can guarantee the victory, cases where both players can win depending on the initial conditions, and finally cases with regions where neither player has a guaranteed victory, and the result depends on the particular players' actions.

The structure of this manuscript is as follows. In Sec.~\ref{sec:Game}, we explore the specifics of the game dynamics. In Sec.~\ref{sec:Solve}, we explain how to resolve it by introducing the algorithm to compute the winning sets. Next, in Sec.~\ref{sec:LogisticMap}, we present a particular game in which one player aims to stay in a chaotic transient region of the logistic map, and the other aims to expel the trajectory from the region. We also discuss the main results. Finally, we present our conclusions.

\section{The Game}
\label{sec:Game}

Imagine two players pulling an object. Player $A$ wants to pull it to its side, while Player $B$ pulls toward its own. Without any other factors involved, the stronger player would simply win by pulling harder.

\begin{figure}
    \centering
    \includegraphics[width=0.7\textwidth]{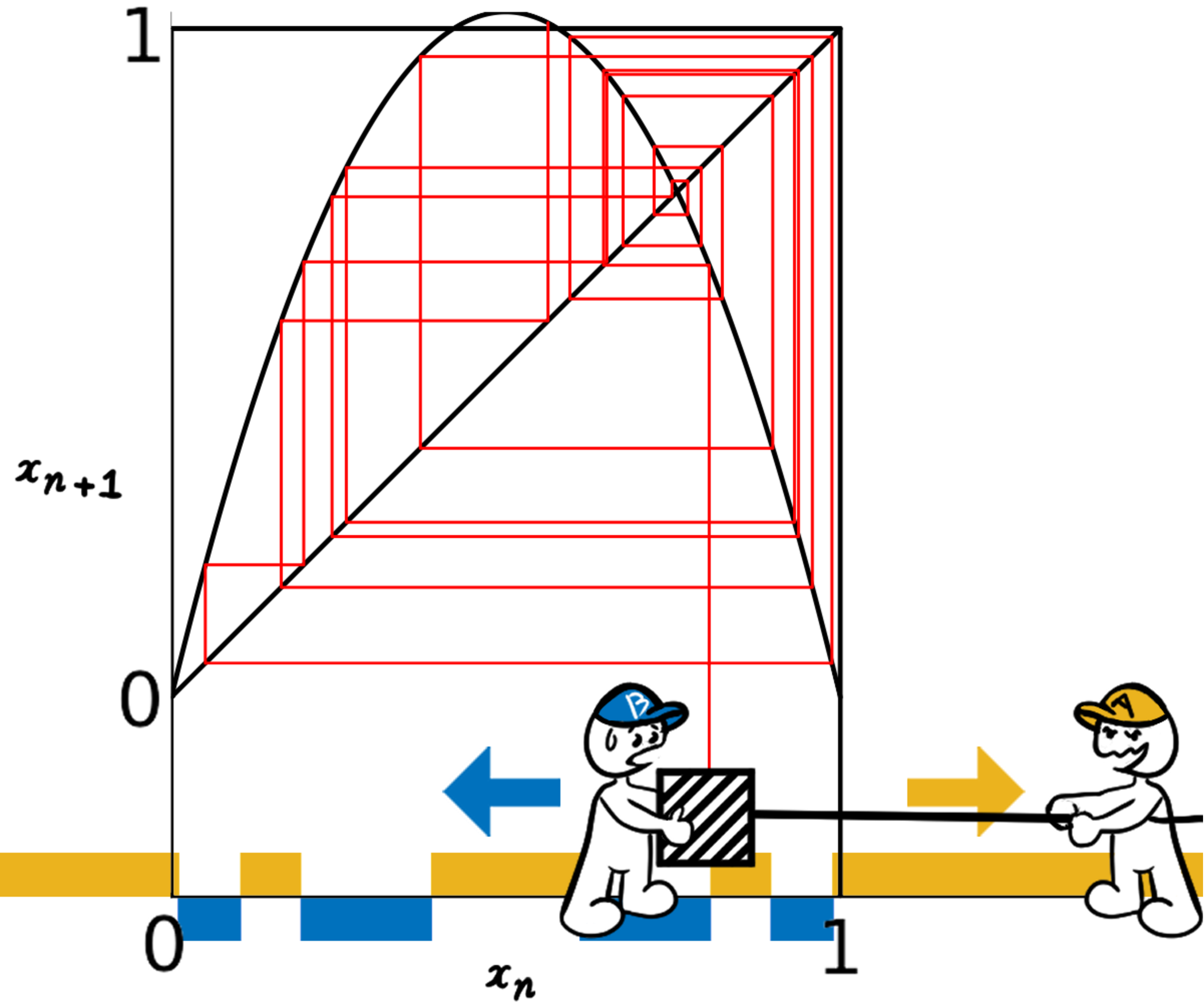}
    \caption{Players $A$ and $B$ compete for controlling the trajectory on the chaotic and escaping logistic map $x_{n+1} = \mu x_n(1-x_n)$. Player $B$ tries to maintain the trajectory on the region $Q = [0,1]$, while fighting against the dynamics of the map, since for $\mu > 4$ all points except a zero measure Cantor set eventually escape from the region. The red line shows an example of one orbit that ends up escaping. Furthermore, player $B$ also fights against the opponent, player $A$, who aims to expel the trajectory from region $Q$ in a finite time. To achieve their goals the players can control the trajectory with a given control bound. The yellow and blue rectangles are the winning sets of players $A$ and $B$, respectively. These are the initial conditions that guarantee victory for each player.}
    \label{fig:drawing}
\end{figure}

However, the game becomes much more interesting when the object naturally moves described by a rule given by a function $f$. Now, both players must not only worry about each other's pulling strength, but also need to consider how the object naturally moves. This becomes especially tricky when the object's movement described by $f$ is chaotic, meaning that even tiny changes in how they pull can lead to completely different and unpredictable results as illustrated in Fig.~\ref{fig:drawing}.

Formally, we describe the game as follows. Consider a region $Q$ of the phase space where a map $f$ acts on the state space $X$. The initial state $x$ of the game starts in $Q$. Player $B$'s objective is to keep the trajectory within $Q$, while player $A$ aims to drive the trajectory outside of $Q$. Each iteration of the game consists of players $A$ and $B$ choosing their respective bounded controls $u_n^A$ and $u_n^B$.  At each discrete time step $n$, the state $x_n$ evolves according to this dynamics
\begin{equation}
    x_{n+1} = f(x_n) + u_n^A + u_n^B,
\end{equation}
where $u_n^A $ and $u_n^B $ represent the control actions of players $A$ and $B$, respectively. These controls are bounded by
\begin{equation}
    |u_n^A| \leq u_0^A, \quad |u_n^B| \leq u_0^B.
\end{equation}

Player $B$ needs to maintain the trajectory inside $Q$ forever to win, while $A$ only needs to drive the trajectory outside $Q$ once to achieve victory.

A critical aspect to solve this game is the order of play. For this reason, we consider three different scenarios:
\begin{table}[h!]
\centering
\begin{tabular}{|p{3cm}|p{4cm}|p{7.3cm}|}
\hline
\textbf{Game Type} & \textbf{Order of Play} & \textbf{Information Available} \\
\hline
Game $A^{-}B^{+}$ & $B$ plays after $A$ & B knows $A$'s action  \\
\hline
Game $A^{+}B^{-}$ & $A$ plays after $B$ & $A$ knows $B$'s action  \\
\hline
Game $A^{-}B^{-}$ & Simultaneous play & Both ignore the opponent's action  \\
\hline
\end{tabular}
\caption{Three different scenarios analyzed in this paper based on the order of play, i.e., the information is available to each player.}
\label{tab:games}
\end{table}

In game $A^{-}B^{+}$, player $B$ has the advantage of knowing player $A$'s action before making their move. This sequential play allows $B$ to react optimally to $A$'s choice. In game $A^{+}B^{-}$, the roles are reversed, with $A$ having complete information about $B$'s move before acting. Finally, in game $A^{-}B^{-}$, both players must make their decisions simultaneously, where both players ignore the action of their opponent.

\section{Solving the game: the winning sets}
\label{sec:Solve}

\begin{figure}[h!]
    \centering
    \includegraphics[trim={0.8cm 0cm 0cm 0cm}, clip,width=0.95\textwidth ]{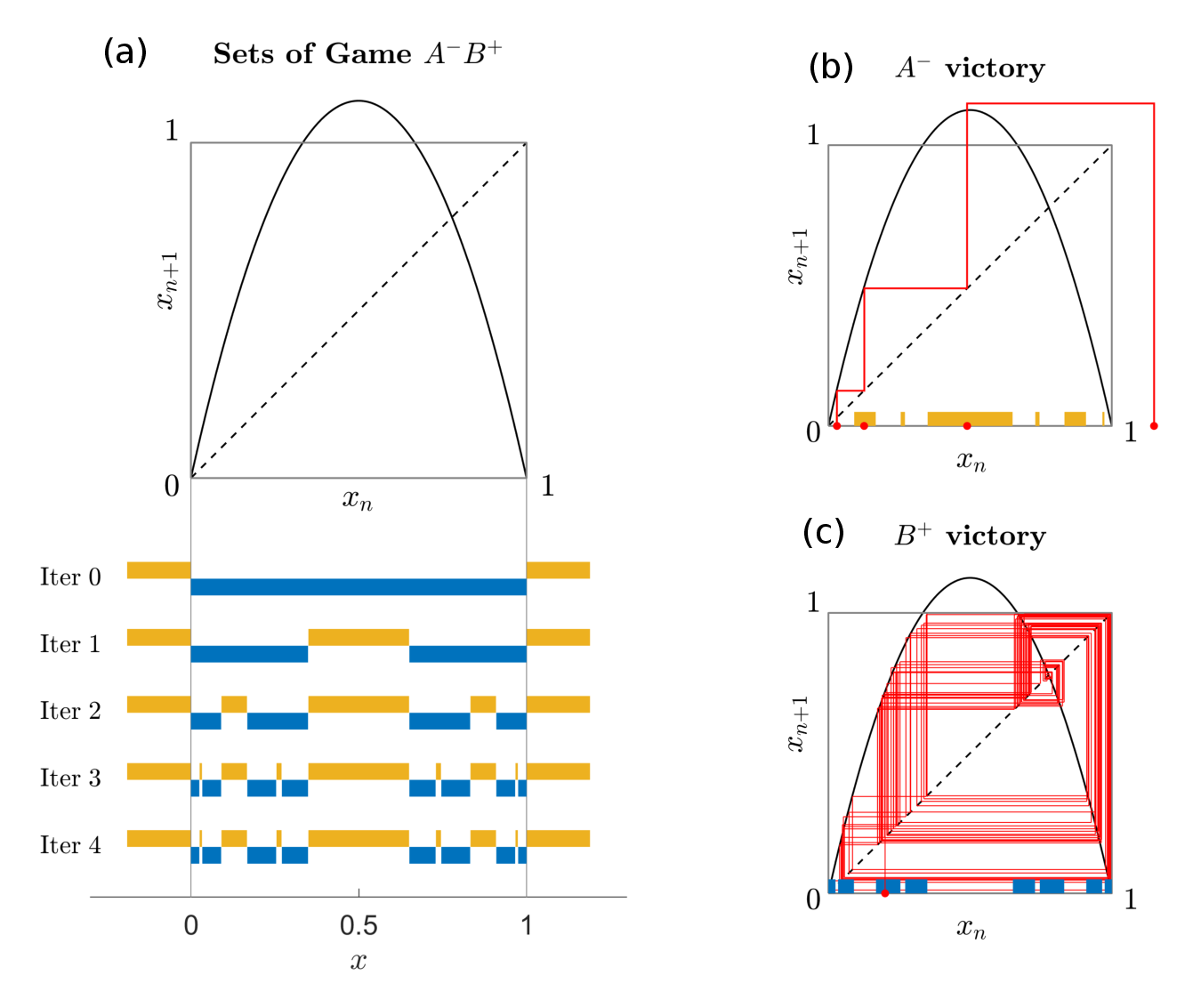}
    \caption{ Winning sets and controlled trajectory for a game between an ignorant player $A$ and an informed player $B$. (a) Steps of the algorithm to compute the winning sets for the ignorant player $A$, $W^{A^-}$, in yellow, and for the informed player $B$, $W^{B^+}$, in blue. The sets were computed in the region $R\in [-0.3,1.3]$, with $Q\in[0,1]$, $u_0^A=0.016$, $u_0^B=0.038$. The algorithm converges in $3$ iterations since  the sets for the third iteration are identical to those of the forth. (b) Controlled trajectory of the game when the initial condition belongs to player's $A$ winning set. Player $A$ acts first, without knowledge of player $B$'s move, so it chooses its control at each step to reach the closest point belonging to the shrunk set ($W^{A^-} -u^B_0$), accounting for the worst possible subsequent action of player $B$. After $3$ iterations the trajectory has left region $Q$, so player $A$ wins. (c) The game now starts in a point that belongs to player's $B$ winning set. Player $B$ acts second, with the knowledge of player $A$'s move, so selects its control to reach the closest point belonging to set $W^{B^+}$. The trajectory stays indefinitely inside region $Q$, so player $B$ wins as long as control is maintained.}
    \label{fig:tray}
\end{figure}

\begin{figure}[h!]
    \centering
   \includegraphics[trim={3.2cm 0cm 0cm 0cm}, clip,width=1.1\textwidth ]{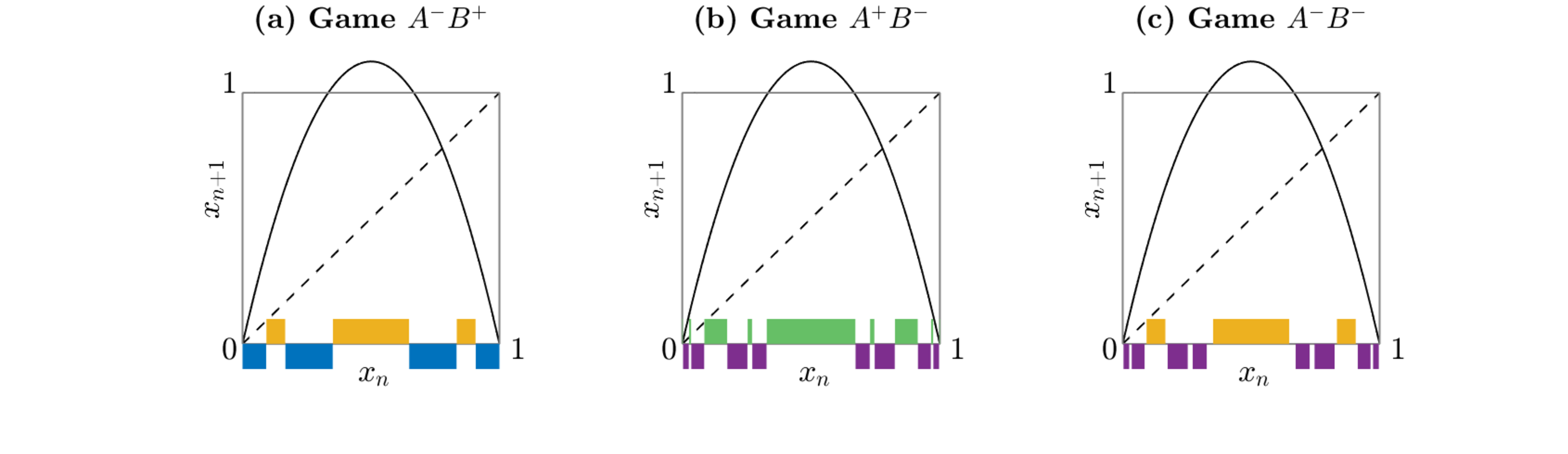}
    \caption{The logistic map and winning sets to control trajectories of the logistic map with $\mu = 4.5$ and control bounds $u^A_0 = 0.015$ for player $A$, and $u^B_0 = 0.040$ for player $B$. Each winning set at the bottom show the initial conditions that guarantee victory to each player. Each player has two winning sets whether they are informed ($+$) or ignorant ($-$) to their opponent's move. The winning sets of player $A$ are colored in green when it is informed and in yellow when it is ignorant. On the other hand, those of player $B$ are colored in blue when it is informed and purple when it is ignorant. For simplicity, we only represent the winning sets in the interval $Q$. (a) Ignorant player $A$ against informed player $B$ (b) Now the informed player is $A$ who plays against the ignorant player $B$ (c) Neither player knows each other's actions. In this case, there are regions that do not belong to any player's winning set, meaning that both player could win when starting at those initial conditions, but the victory is uncertain because players cannot guarantee victory against any move from the opponent. These "no winning regions" represent the difference between the respective informed winning sets and the ignorant winning sets.}
    \label{fig::sets}
\end{figure}

Given the opposing objectives where player $A$ aims to drive trajectories outside $Q$ while player $B$ seeks to keep them within $Q$, we establish a region $R$ containing $Q$ where the winning sets $W$ are defined for both players. These sets have binary values representing whether that initial condition guarantees victory, $1$, or does not, $0$. For player $B$, initially $W^B(x)=1$ for $x \in Q$ and $0$ otherwise. For player $A$, initially $W^A(x)=0$ for $x \in Q$ and $1$ otherwise. The computation procedure for each winning set is described next, which is also summarized in Table~\ref{tab:morphological_operations}. Each set for an ignorant or informed player $A$ or $B$ can be computed independently.

\begin{itemize}
\item {\bf Player $A^-$ (ignorant player $A$):}
To compute the set of $W^{A^-}$, we have to evaluate for each $x \in Q$ the image $f(x)+u^A+u_0^B$ for all possible controls $u^A \in [-u_0^A,u_0^A]$ and $u^B \in [-u_0^B,u_0^B]$. We select only those $u^A$ where all possible images $f(x)+u^A+[-u_0^B,u_0^B]$ fall within $W^{A^-}$. Points $x \in Q$ satisfying this condition update $W^{A^-}(x)=1$. All these initial conditions are able to be controlled to escape after one iteration of the map. But there will be other points that, after more than one iterations of the map and control, will be able to escape. To find this points we must repeat the algorithm, thus updating $W^{A^-}$ until it converges.

This process is equivalent to iteratively performing the following morphological operations: first shrinking $W^{A^-}$ by $u_0^B$, then dilating by $u_0^A$. For any $x \in Q$ whose $f(x)$ falls within the shrunk set, we set $W^{A^-}(x)=1$. The process is repeated with this new $W^{A^-}$ until it converges.

\item {\bf Player $A^+$ (informed player $A$):}
To compute the set of $W^{A^+}$, we analyze $f(x)+u^B+u^A$ for all possible controls. For each $x \in Q$, we evaluate $f(x)+[-u_0^B,u_0^B]+u^A$ and determine if there exists some $u^A \in [-u_0^A,u_0^A]$ capable of forcing escape for all possible images. If such $u^A$ exists, then $W^{A^+}(x)=1$. This process is performed for all $x \in Q$, updating the set $W^{A^+}$. The algorithm is repeated until $W^{A^+}$ converges.

Morphologically, this process is equivalent to first dilating $W^{A^+}$ with $u_0^A$ and then shrinking the resulting set with $u_0^B$. For any $x \in Q$ whose $f(x)$ falls within this shrunk set, we set $W^{A^+}(x)=1$.

\item {\bf Player $B^-$ (ignorant player $B$):}
To compute the set of $W^{B^-}$, we evaluate for each $x \in Q$ the image $f(x)+u^B+u_0^A$ for all possible controls $u^B \in [-u_0^B,u_0^B]$ and $u^A \in [-u_0^A,u_0^A]$. We select only those $u^B$ where all possible images $f(x)+u^B+[-u_0^A,u_0^A]$ fall within $W^{B^-}$. Points $x \in Q$ not satisfying this condition update $W^{B^-}(x)=0$. Since player $B$ wants to keep the orbit in $Q$ forever, the algorithm must be repeated until $W^{B^-}$ converges.

The process is equivalent to iteratively performing the following morphological operations: first shrinking $W^{B^-}$ by $u_0^A$, then dilating by $u_0^B$. For any $x \in Q$ whose $f(x)$ falls outside the shrunk set, we set $W^{B^-}(x)=0$. The process is repeated with this new $W^{B^-}$ until it converges.

\item {\bf Player $B^+$ (informed player $B$):}
To compute the set of $W^{B^+}$, we analyze $f(x)+u^A+u^B$ for all possible controls. For each $x \in Q$, we evaluate $f(x)+[-u_0^A,u_0^A]+u^B$ and determine if there exists some $u^B \in [-u_0^B,u_0^B]$ capable of guaranteeing containment for all possible images. If such $u^B$ does not exist, then $W^{B^+}(x)=0$. This process is performed for all $x \in Q$, updating the set $W^{B^+}$. The algorithm is repeated until $W^{B^+}$ converges.

Morphologically, this process is equivalent to first dilating $W^{B^+}$ with $u_0^B$ and then shrinking the resulting set with $u_0^A$. For any $x \in Q$ whose $f(x)$ falls outside this shrunk set, we set $W^{B^+}(x)=0$.
\end{itemize}

\begin{table}[h]
\centering
\begin{tabular}{|c|c|>{\raggedright\arraybackslash}p{10.2cm}|}
\hline
Player & Initial Set & \multicolumn{1}{c|}{Morphological Operations} \\
\hline
$A^-$ & 
$\begin{array}{l} 
W^{A^-}(x)=\begin{cases} 
0 & x \in Q \\
1 & \text{otherwise}
\end{cases} \\
\\
W^{A^-}_{\text{new}}=W^{A^-}
\end{array}$ 
& $\begin{array}{l} \\ 
1. \text{ Shrink } W^{A^-} \text{ by } u_0^B \text{ to obtain } W^{A^-}_{\text{shrunk}} \\ 2. \text{ Dilate } W^{A^-}_{\text{shrunk}} \text{ by } u_0^A \text{ to obtain } W^{A^-}_{\text{dilated}} \\ 3. \text{~} \forall x \in Q, \text{ if } f(x) \text{ falls in } W^{A^-}_{\text{dilated}},\text{ set }W^{A^-}_{\text{new}}(x)=1 \\  4. \text{~} W^{A^-}=W^{A^-}_{\text{new}}. \text{ Go to step 1 and repeat the process} \\ ~
\end{array}$ \\ \hline
$A^+$ & $\begin{array}{l}
W^{A^+}(x)=\begin{cases}
0 & x \in Q \\
1 & \text{otherwise}
\end{cases} \\
\\
W^{A^+}_{\text{new}}=W^{A^+}
\end{array}$ 
& $\begin{array}{l} \\ 
1. \text{ Dilate } W^{A^+} \text{ by } u_0^A \text{ to obtain } W^{A^+}_{\text{dilated}} \\ 2. \text{ Shrink } W^{A^+}_{\text{dilated}} \text{ by } u_0^B \text{ to obtain } W^{A^+}_{\text{shrunk}} \\ 3. \text{~} \forall x \in Q, \text{ if } f(x) \text{ falls in } W^{A^+}_{\text{shrunk}},\text{ set }W^{A^+}_{\text{new}}(x)=1 \\  4. \text{~} W^{A^+}=W^{A^+}_{\text{new}}. \text{ Go to step 1 and repeat the process} \\ ~
\end{array}$ \\ \hline
$B^-$ & $\begin{array}{l}
W^{B^-}(x)=\begin{cases}
1 & x \in Q \\
0 & \text{otherwise}
\end{cases} \\
\\
W^{B^-}_{\text{new}}=W^{B^-}
\end{array}$ 
& $\begin{array}{l} \\ 
1. \text{ Shrink } W^{B^-} \text{ by } u_0^A \text{ to obtain } W^{B^-}_{\text{shrunk}} \\ 2. \text{ Dilate } W^{B^-}_{\text{shrunk}} \text{ by } u_0^B \text{ to obtain } W^{B^-}_{\text{dilated}} \\ 3. \text{~} \forall x \in Q, \text{ if } f(x) \text{ falls in } W^{B^-}_{\text{dilated}},\text{ set }W^{B^-}_{\text{new}}(x)=1 \\  4. \text{~} W^{B^-}=W^{B^-}_{\text{new}}. \text{ Go to step 1 and repeat the process} \\ ~
\end{array}$ \\ \hline
$B^+$ & $\begin{array}{l}
W^{B^+}(x)=\begin{cases}
1 & x \in Q \\
0 & \text{otherwise}
\end{cases} \\
\\
W^{B^+}_{\text{new}}=W^{B^+}
\end{array}$ 
& $\begin{array}{l} \\ 
1. \text{ Dilate } W^{B^+} \text{ by } u_0^B \text{ to obtain } W^{B^+}_{\text{dilated}} \\ 2. \text{ Shrink } W^{A^+}_{\text{dilated}} \text{ by } u_0^A \text{ to obtain } W^{B^+}_{\text{shrunk}} \\ 3. \text{~} \forall x \in Q, \text{ if } f(x) \text{ falls in } W^{B^+}_{\text{shrunk}},\text{ set }W^{B^+}_{\text{new}}(x)=1 \\  4. \text{~} W^{B^+}=W^{B^+}_{\text{new}}. \text{ Go to step 1 and repeat the process} \\ ~
\end{array}$ \\ \hline
\end{tabular}
\caption{Morphological operations for each player's winning set computation.}
\label{tab:morphological_operations}
\end{table}


\section{Application of the  Game in the Logistic Map}
\label{sec:LogisticMap}

To demonstrate our approach, we study a game based on the logistic map ${f(x) = \mu x_n(1-x_n)}$ where $\mu=4.5$ and $Q=[0,1]$. This map serves as an excellent test case since it exhibits transient chaos in $Q$: trajectories display chaotic behavior before naturally escaping the region. This creates an asymmetric scenario where the player $B$, who wants to stay in the region, must work against both the natural dynamics and their opponent's actions, who has an opposite objective, i.e., to expel the trajectory from the region. While this specific example is used, our methodology can be applied to others maps $f$ and defined regions $Q$ in phase space.

In this context, both players can apply control at each iteration to influence the trajectory, but they are limited by their respective control bounds $u_0^A$ and $u_0^B$. The state of the system at each iteration is determined by
\begin{equation}
x_{k+1} = f(x_k) + u^A + u^B,
\end{equation}
where $f(x) = \mu x(1-x)$ is the logistic map, and $u^A$ and $u^B$ represent the control inputs applied at each iteration according to each player's strategy.

We compute an example for this game with specific control bounds $u_0^A = 0.016$ and $u_0^B = 0.038$. In Fig.~\ref{fig:tray}, we show how the winning sets are constructed. Each iteration of the process brings the winning sets closer to their final version by cutting down player's $B$ set and expanding that of player $A$. Additionally, we show two controlled trajectories with initial conditions belonging to the winning set of player $A$ in Fig.~\ref{fig:tray}(b), successfully expelling the trajectory from region $Q$; and of player $B$ in Fig.~\ref{fig:tray}(c), keeping the orbit in the region. 

Figure~\ref{fig::sets} shows three games wether each player is informed or ignorant, showing the winning sets of each player. In Fig.~\ref{fig::sets}(a), we analyze the case where $A$ plays first and $B$ second. In Fig.~\ref{fig::sets}(b), the roles are reversed: $B$ plays and $A$ plays second. Finally, Fig.~\ref{fig::sets}(c) shows the scenario where both players are ignorant, which could represent a simultaneous play situation.

Particularly interesting is the case shown in Fig.~\ref{fig::sets}(c), where both players play simultaneously and therefore they are ignorant of each other's actions. Here, we observe regions in the state space that do not belong to any winning set. These regions emerge because neither player can guarantee victory regardless of their strategy. The existence of such regions is a direct consequence of both players being ignorant, no one knows the other player's move when making their decision. Therefore, players may or may not choose the best response to the opponent's move so the outcome depends on the specific combination of controls chosen by both players

This uncertainty contrasts sharply with the scenarios where at least one player is informed, as shown in Fig.~\ref{fig::sets}(a-b), where the winning sets completely determine the game's outcome for all initial conditions.

\subsection{Exploring possible games in the $(u_0^B,u_0^A)$ space}
\begin{figure}
    \centering
    \includegraphics[trim={0.6cm 0cm 0cm 0cm}, clip,width=1.12\textwidth ]{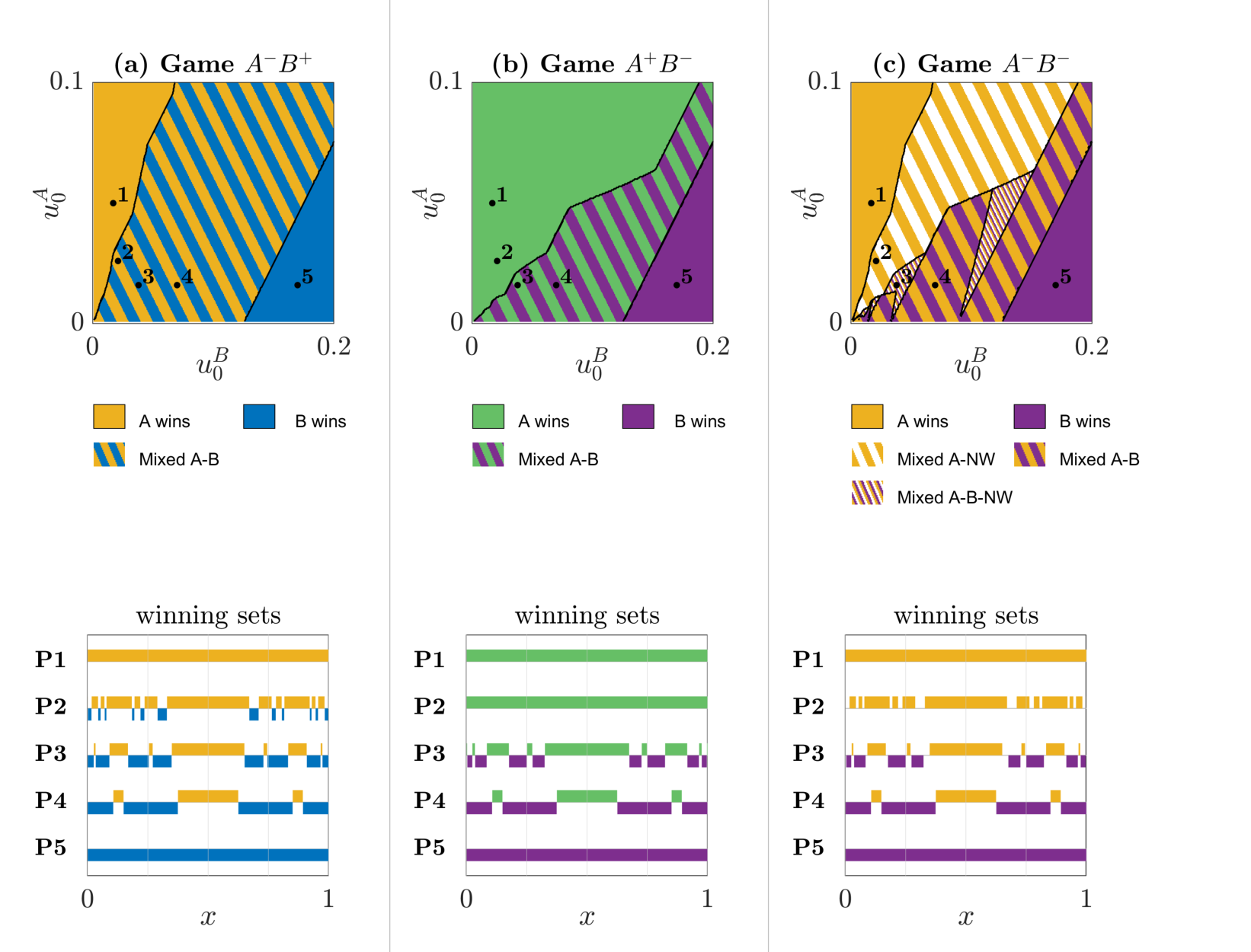}
    \caption{Different solutions depending on the values of $u^A_0$ and $u^B_0$. Panels (a), (b), and (c) show a diagram in which solid colored regions represent parameter combinations where the informed/ignorant player $A$ wins for all initial conditions at solid colors green/yellow and the informed/ignorant player $B$ at solid blue/purple. Striped regions indicate parameters where the winner depends on the initial conditions. There are also regions with white strips marked as $NW$ (no winning), where the victory is uncertain at some initial conditions. Below each diagram we show the winning sets, i.e., the initial conditions that guarantee victory for each players, for each kind of game at $5$ different regions of control bound. These points, written as $(u^B_0, u^A_0)$, are: $P1 =  (0.017, 0.050)$, $P2 =  (0.021, 0.026)$, $P3 =  (0.038, 0.016)$, $P4=  (0.070, 0.016)$, and $P5= (0.170, 0.016)$. 
    }
    \label{fig:franjas}
\end{figure}

\begin{figure}
    \centering
    \includegraphics[trim={3.1cm 0cm 0cm 0cm}, clip,width=1.12\textwidth  ]{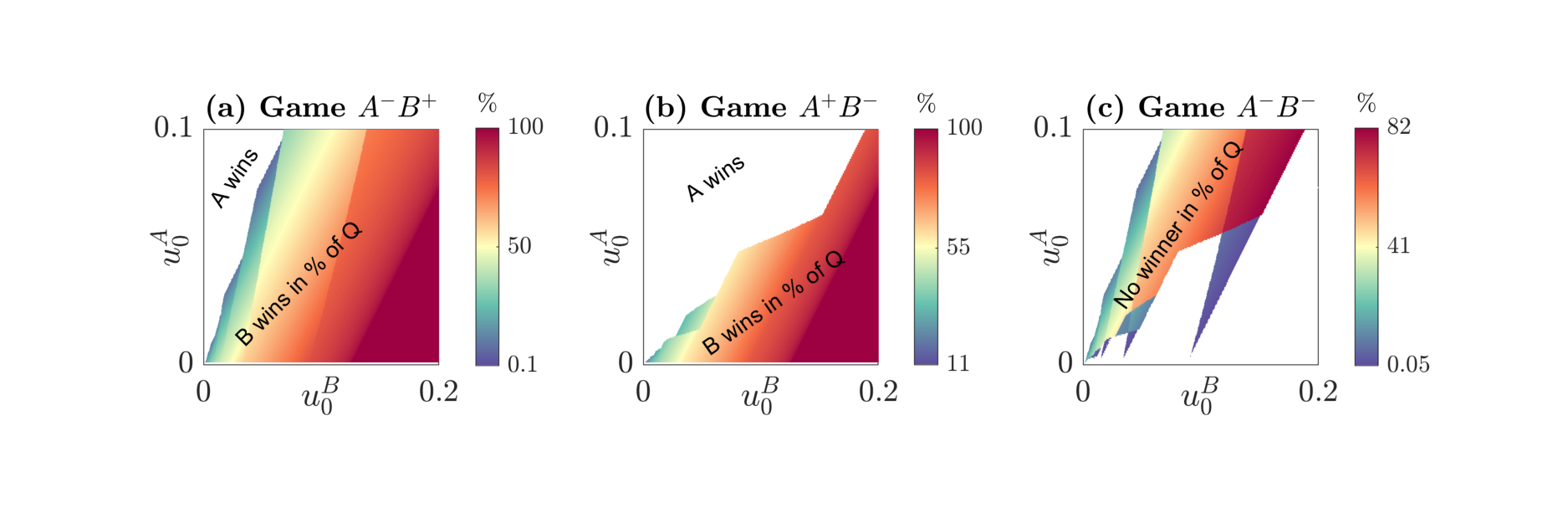}
    \caption{The first two panels show the percentage of the region that occupies the winning set of player $B$ respect region $Q$ whether it is informed (+) or ignorant (-). The color is white when the percentage is zero. The last panel represents the percentage of initial conditions that do not assure a winner when both players are ignorant. All figures show similar details at different scales, suggesting self similarity in the system dynamics.}
    \label{fig:fraccion}
\end{figure}

The outcome of these games strongly depends on the control bounds available to each player. Figure~\ref{fig:franjas} presents a comprehensive analysis of possible game scenarios in the $(u_0^B,u_0^A) \in [0,0.2]\times[0,0.1]$ parameter space. It shows three different game types: $A$ ignorant vs $B$ informed, $A$ informed vs $B$ ignorant, and both players ignorant.

In these diagrams, there are regions where one player wins for all initial conditions and other regions in which the winner depends on the initial conditions. Among this last category, in the case where both players are ignorant in Fig.~\ref{fig:franjas}(c), some are marked as $NW$ (no winning) where victory is uncertain for some initial conditions. Five different regions can be seen in Fig.~\ref{fig:franjas}(c). This variety of outcomes illustrates the complexity that emerges when both players must act without knowledge of their opponent's moves.

The bottom Figs.~\ref{fig:franjas}(d-f) illustrate specific examples of winning sets for five different parameter combinations, labeled as games $P1$ to $P5$. These examples demonstrate how the winning sets change as we move through different regions in the parameter space.

Figure~\ref{fig:fraccion} quantifies these results by showing the percentage of region $Q$ occupied by each player's winning set. Particularly interesting is Fig.~\ref{fig:fraccion}(c), which shows the percentage of initial conditions where no player can guarantee victory in the case where both players are ignorant. The color gradients reveal a continuous transition between different game outcomes as the control bounds change.

These results demonstrate how the relative strength of the players' control bounds determines not just who wins, but whether victory can be guaranteed at all. When the bounds are comparable, we often find situations where the outcome depends on initial conditions or remains uncertain.

\subsection{Boundary games}

\begin{figure}
    \centering
    \includegraphics[trim={3.3cm 0cm 0cm 0cm}, clip,width=1.06\textwidth ]{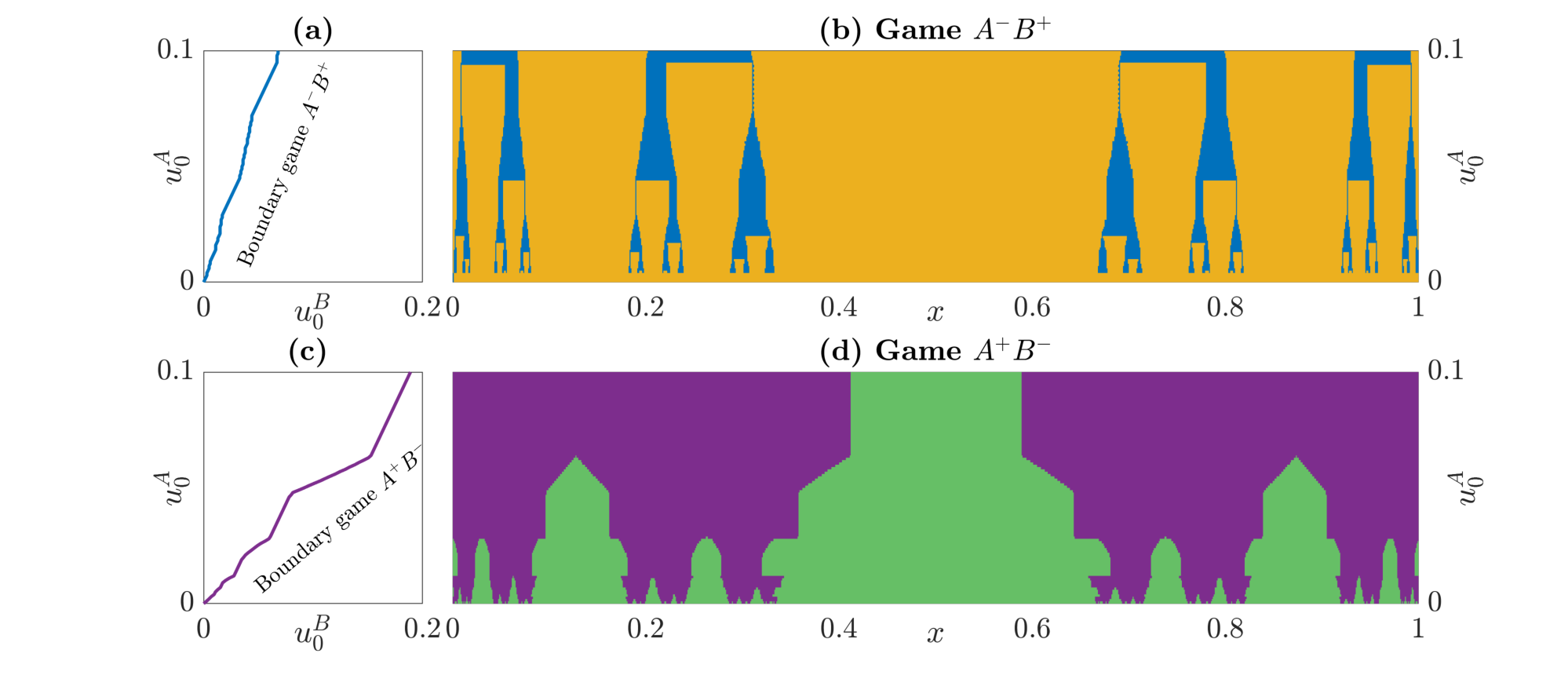}
    \caption{Panels (a) and (c) show the game boundary for which there is a shift between player $A$ winning the game at all initial conditions (left of the line) and there is a mixed victory among different initial conditions (right of the line). This lines are the boundaries of panels (a) and (b) from the previous figure. Panels (b) and (d) show the winning sets within the control bounds along the game boundary. In colors yellow and blue the points in the $A$ ignorant and $B$ informed winning sets respectively and in green and purple, $A$ informed and $B$ ignorant. In panel (b) we can clearly see increasing and almost self-similar details when the scales decreases.}
    \label{fig:bifurcation}
\end{figure}

The dynamics of the game favors player $A$, who wants to leave region $Q$, since this occurs spontaneously due to the transient nature of the system. Nonetheless all is not lost for player $B$. Given sufficient control this player can achieve its goal. The boundary games illustrated in Fig.~\ref{fig:bifurcation} explore the limits of control when player $B$'s  victories first appear. This boundary, shown in Figs.~\ref{fig:bifurcation}(a) and~\ref{fig:bifurcation}(c) in the $(u_0^B,u_0^A)$ parameter space, represents a critical transition: for slightly larger values of $u_0^A$ or smaller values of $u_0^B$, player $B$ cannot win under any circumstances. However, crossing this boundary leads to the emergence of initial conditions where $B$ can achieve victory.

Of particular interest is the structure revealed in Figs.~\ref{fig:bifurcation}(b) and~\ref{fig:bifurcation}(d), which show the winning sets along these boundaries for two different game scenarios. In the case of $A$ ignorant vs $B$ informed, Fig.~\ref{fig:bifurcation}(b), we observe an intricate pattern of $B$'s victories appearing as blue regions within $A$'s domain, showing increasing detail at smaller scales and suggesting a self-similar structure. In contrast, when $A$ is informed and $B$ ignorant, Fig.~\ref{fig:bifurcation}(d), we see a different pattern where $B$'s victory regions (purple) alternate with $A$'s (green) in a complex but distinct arrangement.

\subsection{Is it worth being informed?}

\begin{figure}
    \centering
    \includegraphics[trim={0cm 0cm 0cm 0cm}, clip,width=0.5\textwidth ]{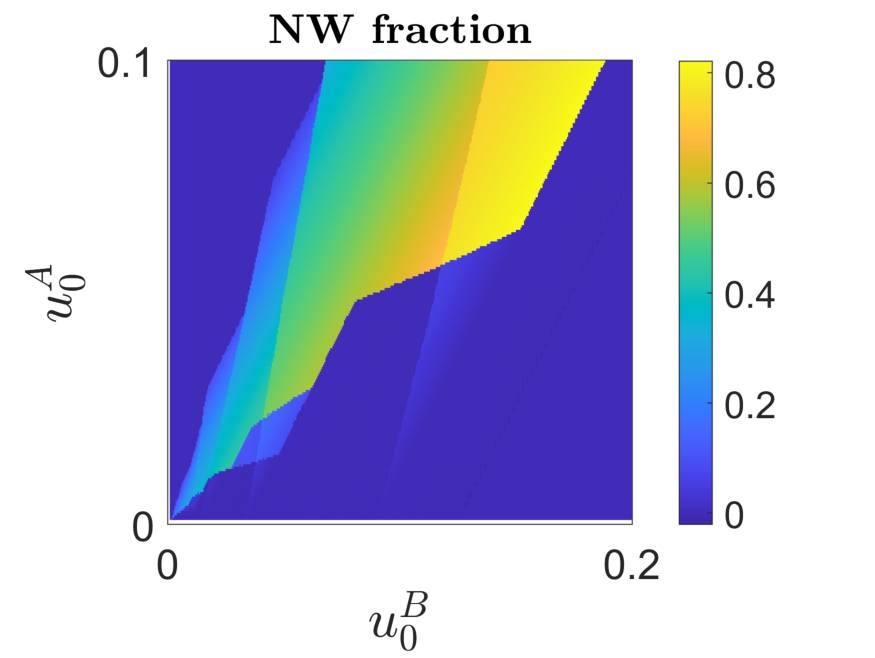}
    \caption{The color on the heatmap represents the Difference between the informed and ignorant sets fraction. The no winning fraction ($NW$) satisfies the relation $NW = A^{+} - A^{-} = B^{+} - B^{-}$, that is, the difference is the same for both players. Therefore they will fight for being the one informed at the same parameter regions. However for many points the difference is $0$, so be informed or be ignorant results in the same winning set.}
    \label{fig:diferencia}
\end{figure}

It is clear that the second player has advantage in this game if it has the knowledge of the opponent's actions because it can counteract any action of the rival. Consequently, the ignorant sets are a subset of the informed ones. But, surprisingly, by plotting the difference between the proportion of each player informed and ignorant sets, Fig.~\ref{fig:diferencia}, we can see that there is a major area where the difference is null. This figure is the same as the percentage in which no player guarantees its victory of Fig.~\ref{fig:fraccion}. This is easily demonstrated with the following equations:
\begin{equation}
   \left. \begin{array}{lll}
        &A^{+} + B^{-} = 1 \\
        &A^{-} + B^{+} = 1 \\
        &A^{-} + B^{-} + NW = 1
    \end{array}
    \right\} \Rightarrow NW = A^{+} - A^{-} = B^{+} - B^{-}.
\end{equation}
Here $P^{+}$ represents the proportion of the region $Q$ that is occupied by the informed Player $P$'s winning set while $P^{-}$ the same for the ignorant one. $ NW$ (no winning) is the proportion of blank space when both Player $A$ and Player $B$ are ignorant.

In the case that players could fight for turn order, it is reasonable to assume that there would be a cost associated to being informed. If that is the case, then it may not be worth paying that cost, because the result is the same for a large proportion of cases. However, if one cares to minimize the average control made, being informed is surely beneficial in almost all cases.

\section{Conclusions}

This work has analyzed a novel game of survival in a chaotic region, where players have opposing objectives: one aims to maintain the system within a region while the other attempts to escape it. The game was solved in all situations with the help of the partial control method. This method provides explicit solutions for determining victory conditions, allowing us to construct winning sets for each player under different information scenarios. The solutions are not unique, since the partial control method does not provide a unique trajectory but a range of possible points that the trajectory must not leave, the winning sets.

By mapping the complete $(u_0^B,u_0^A)$ parameter space, we identified regions where one player has complete victory, other regions where victory depends on initial conditions, and even regions with initial conditions that does not assure victory for neither player in the cases of mutual ignorance.

The game reveals intricate, self-similar structures in the transition regions where $B$'s victories first emerge, suggesting fractal-like properties in the system's dynamics.

A particularly interesting finding is that, in some cases, identical winning sets are obtained through the informed and the ignorant scenarios. Nonetheless, control with less average control can be achieved by being informed, though the winning conditions do not vary. 

Also interesting is the fact that the game is not solved entirely in the case where both players are ignorant to each other's actions. For some initial conditions, the victory is not assured for neither player. Therefore, victory depends on both player's actions.

These results not only advance our understanding of chaotic control in competitive scenarios, but also suggest practical implications for control strategy design when complete information is not available.

\section*{Acknowledgments}

This work has been supported by the Spanish State Research Agency (AEI) and the European Regional Development Fund (ERDF, EU) under Project No.~PID2023-148160NB-I00 (MCIN/AEI/10.13039/ 501100011033).

\end{document}